\documentclass[superscriptaddress,prl,floatfix,twocolumn]{revtex4-1}
\usepackage{graphicx}
\usepackage{amsmath}
\usepackage{bm}
\usepackage{color}

\begin{document}

\title{Pressure-induced structural transformation of clathrate Ge$_{136}$ via an ultrafast recrystallization of an amorphous intermediate}

\author{Mari\'{a}n Ryn\'{i}k}
\email{marian.rynik@fmph.uniba.sk}
\affiliation{Department of Experimental Physics, Comenius University, Mlynsk\'{a} Dolina F2, 842 48 Bratislava, Slovakia}

\author{Stefano Leoni}
\email{leonis@cardiff.ac.uk}
\affiliation{School of Chemistry, Cardiff University, Cardiff, CF10 3AT, UK}

\author{Roman Marto\v{n}\'{a}k}
\email{martonak@fmph.uniba.sk}
\affiliation{Department of Experimental Physics, Comenius University, Mlynsk\'{a} Dolina F2, 842 48 Bratislava, Slovakia}

\date{\today}

\begin{abstract}
We study the pressure-induced structural transformation of Ge$_{136}$ clathrate by ab initio molecular dynamics and metadynamics. The system under pressure first undergoes amorphization followed by an ultrafast recrystallization to the $\beta$-tin structure on the time scale of 30 ps. 
The initial pressure-induced  amorphization of clathrate is triggered by high pressure while the subsequent fast recrystallization to $\beta$-tin is driven by low temperature. Interestingly, the amorphous intermediate is still diffusive even at room temperature, in spite of very strong undercooling, making the ultrafast recrystallization possible. The system provides an explicit example of structural transformation between two crystalline phases proceeding via non-crystalline intermediate.
Upon fast decompression of the amorphous structure with incipient crystalline order the recrystallization is blocked and the system instead proceeds to the tetrahedral LDA amorphous phase. 
\end{abstract}

\maketitle

Open framework structures like clathrates~\cite{Cros:1970gy,Kasper:1965kp,SanMiguel:1999ev} are enjoying a renewed interest for they may host distinct electronic, magnetic, spectral and transport properties~\cite{Melinon:1998fy,Demkov:1996gu,Saito:1995ff,Nesper:1993fga,OKeeffe:1992gf}. Due to their low density, clathrates are thermodynamically metastable at ambient conditions and typically become stable only at negative pressures. It can therefore be expected that even upon a modest compression they would readily transform to denser phases. It is less obvious, however, how a complicated order in a large unit cell containing $\sim 10^2$ atoms could transform to standard crystalline phase with few atoms in the unit cell. Certainly, it is difficult to imagine any kind of simple martensitic mechanism directly accomplishing such transformation.

Interesting examples are clathrates of group IV elements, in particular Si and Ge (for review see Ref.\cite{review_clathrates}). Type-II clathrate Ge(cF136) allotrope was synthesised from a salt precursor Na$_{12}$Ge$_{17}$, by mild oxidation with HCl. It is stable at room condition and subsists up to 693 K~\cite{Guloy:2006gi}.
Amorphous minority phases are encountered during Ge$_{136}$ synthesis\cite{Guloy:2006gi}.
Its pressure-induced structural changes and its behaviour upon compression were studied in Ref.~\cite{Schwarz:2008wl} for a sample containing nearly 2 \% of impurities and several transformations into denser crystalline Ge phases (hR8, tI4) were found via compression above 7.6 GPa. No such study is known for guest-free Ge$_{136}$.
In high-pressure studies of guest-free Si$_{136}$ clathrate\cite{Tang2006, Ramachandran_2000} the system was found to transform at pressure of 8-10 GPa to the 6-coordinated $\beta$-tin structure. Thermodynamically, it was found from DFT calculations that the $\beta$-tin structure becomes more stable with respect to the Si$_{136}$ clathrate already at pressure of 3-4 GPa. The need for substantial overpressurization 
was attributed to the absence of a convenient low-energy pathway for the Si$_{136}$ to $\beta$-tin transition. A similar study was performed in Ref.\cite{PhysRevLett.83.5290} where Si$_{34}$ clathrate also transformed to the $\beta$-tin structure at pressure of 11 GPa.

Here we focus on the pressure-induced transformations of Ge$_{136}$ and study them
by means of ab initio simulations. The phase diagrams of Si and Ge are very similar, but the range of stability of the $\beta$-tin phase in Ge is much wider than in Si\cite{Ge_phase_diagram_2021}.
The aim of our study is twofold: find a theoretical prediction for the high-pressure behaviour of Ge$_{136}$ clathrate and uncover the elusive microscopic mechanism of the transformation. We employ ab initio molecular dynamics (MD) and metadynamics \cite{metadynamics, PhysRevLett.90.075503, natmat2006}. 
Ab initio calculations were performed by the VASP code\cite{vasp,vasp_paw} and technical details are described in Supplementary Material (SM). 

The enthalpies of the Ge$_{136}$ clathrate, cubic diamond and $\beta$-tin structure are shown in Fig.1 (SM) where it can be seen that the $\beta$-tin structure becomes more stable with respect to clathrate at 4 GPa.
In order to assess the ultimate structural stability of the clathrate upon compression we started with static compression at T=0, increasing pressure in 5 GPa steps. Clathrate survived up to 30 GPa while at 35 GPa it reached a limit of mechanical stability and quickly transformed into a 6-coordinated disordered structure. The evolution of structure during 200 geometry optimization steps is shown in Fig.2 (SM). Interestingly, in the initial phase of the collapse the structure stratifies into (111) layers of coordination 4 and 5.
The final disordered structure has a radial distribution function (RDF) which is remarkably structureless in the region from 3 to 4.5 $\AA$ (Fig.3a (SM)). The angular distribution (Fig.3b (SM)) has a sharp peak at 60$^\circ$ and for higher angles is nearly flat up to 130$^\circ$ where it starts to drop, also pointing to substantial disorder.

In order to make sure that the pressure-induced amorphization (PIA) observed at $T=0$ is not just an artefact of overpressurization we performed ab initio metadynamics employing the cell as collective variable \cite{natmat2006} at a lower pressure of 10 GPa and $T=300$ K. At finite temperature the system is able to cross the barriers and one can expect the structural transformation to start at lower pressure, possibly resulting in different and more ordered structure. 

\begin{figure}[htpb]
  \includegraphics*[width=\columnwidth]{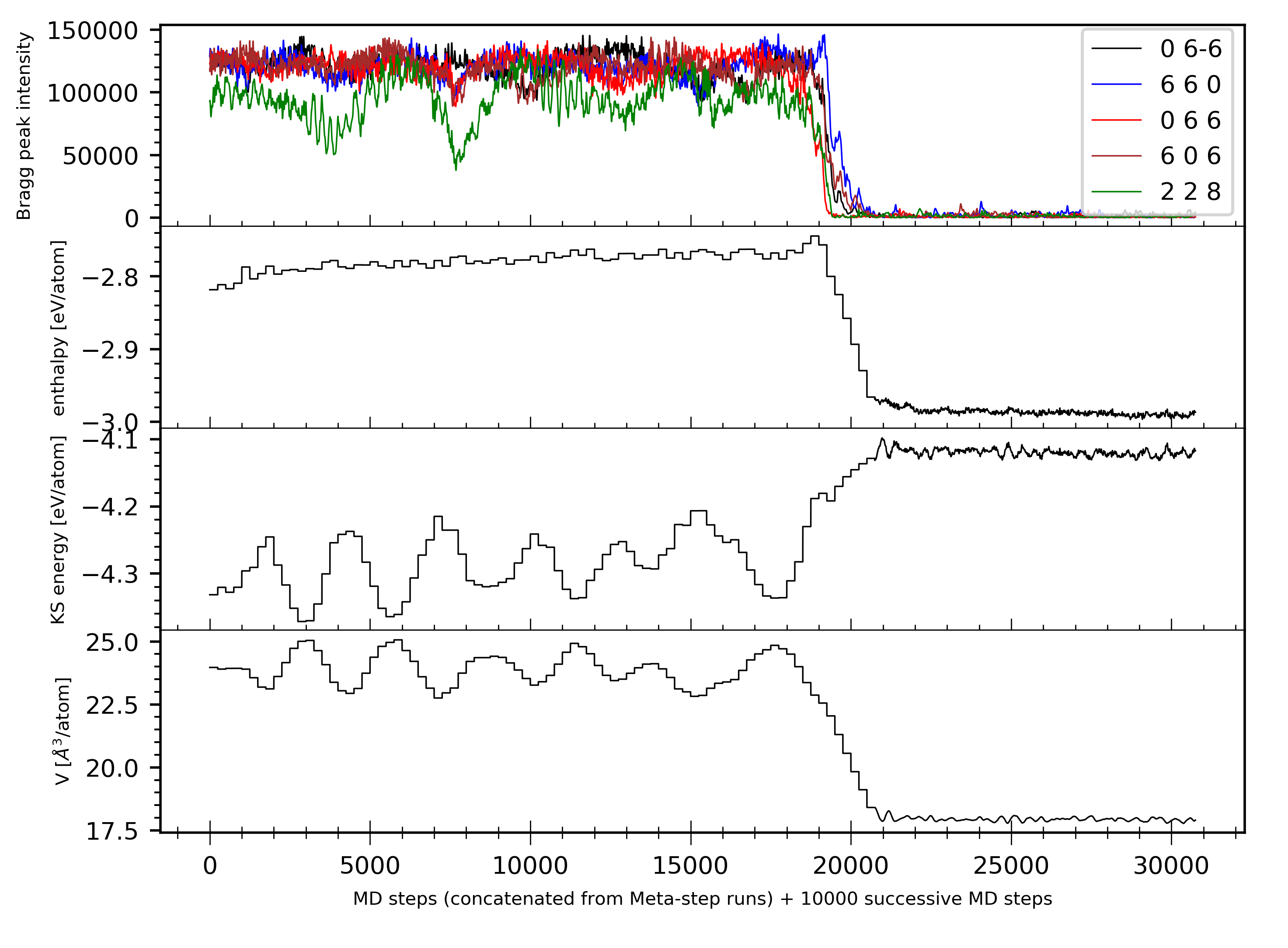}
    \includegraphics*[width=\columnwidth]{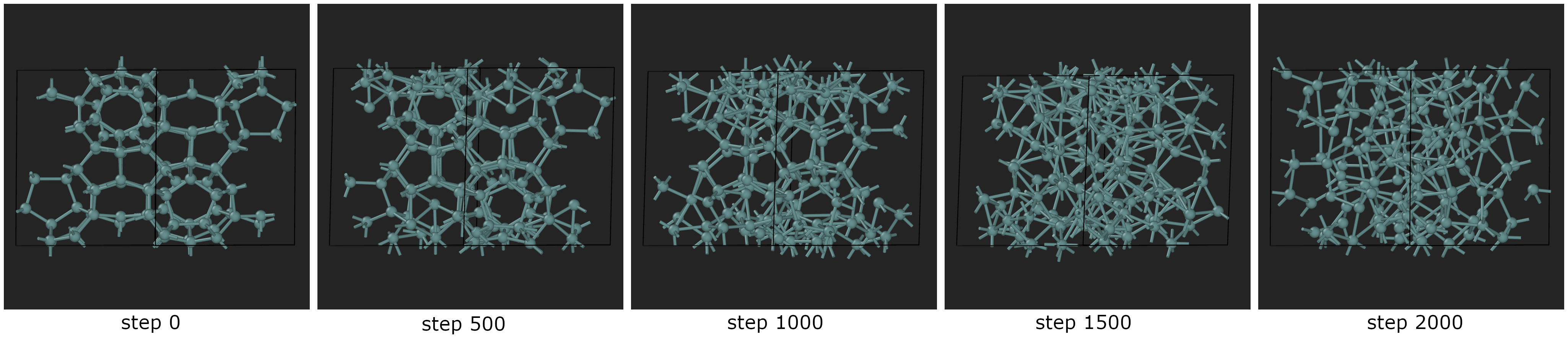}
  \caption{(a) upper panel: Time evolution of volume, energy, enthalpy and intensity of selected Bragg peaks of clathrate structure in metadynamics at $p=100$ kbar and $T = 300$ K, resulting in amorphization of Ge$_{136}$ clathrate after 83 metasteps. (b) lower panel: Sequence of configurations from metadynamics steps 76-83 (2000 MD steps) where pressure-induced amorphization of clathrate occurs. The figure was prepared using the OVITO package\cite{ovito}.}
 \label{fig:clathrate_collapse_100kbar}
\end{figure}

The evolution of structure and energy of the clathrate is shown in Fig.~\ref{fig:clathrate_collapse_100kbar} (a). During the first 76 metasteps (19000 MD steps) the enthalpy of the system gradually grows while the volume, energy and intensity of Bragg peaks oscillate as the system explores the initial free-energy basin.
After 76 metasteps the clathrate structure starts to collapse, signaled by vanishing of the Bragg peaks. The collapse is accomplished at metastep 83, accompanied by large volume drop of about 25 \% and enthalpy drop of 0.16 eV/atom. The collapse again produces a disordered structure (see the structural evolution in Fig.\ref{fig:clathrate_collapse_100kbar} (b)) with very similar RDF (Fig.3 (a)) as the structure produced at $T=0$ (Fig.3a (SM)). Metadynamics simulation of room-temperature compression thus confirms the pressure-induced amorphization (PIA) scenario found at $T=0$.

\begin{figure}[htpb]
  \includegraphics*[width=\columnwidth]{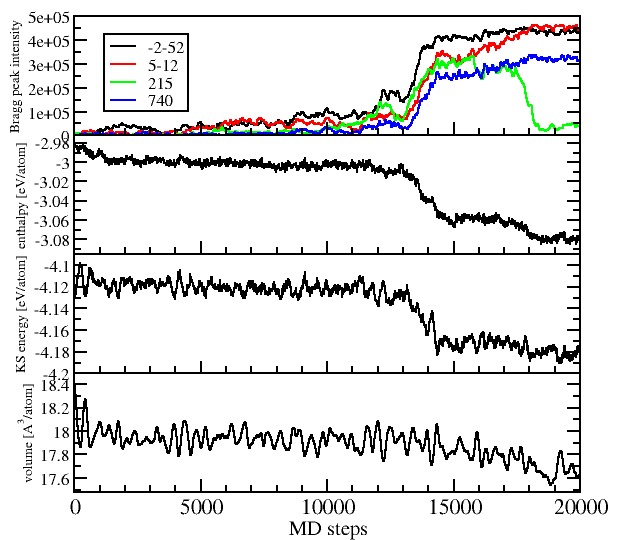}
      \includegraphics*[width=\columnwidth]{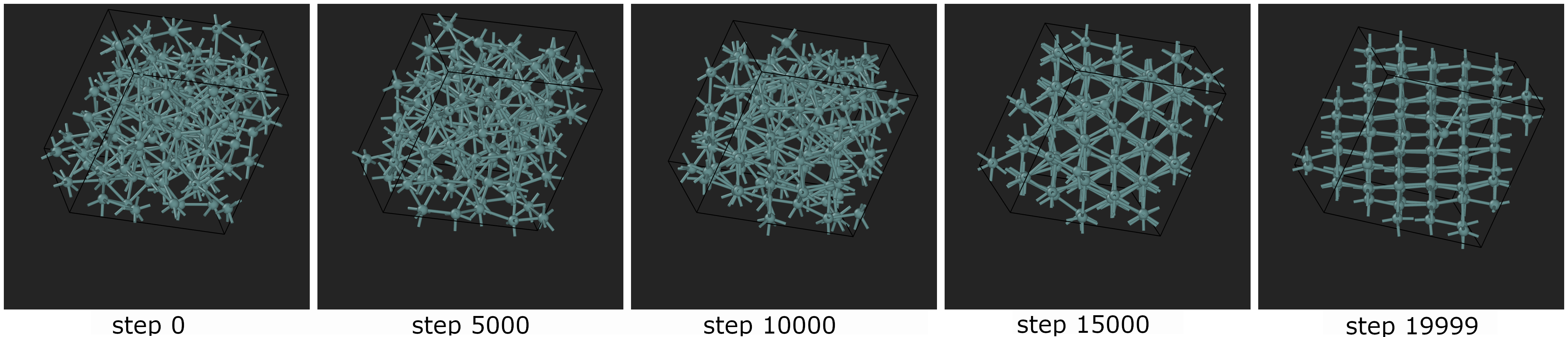}
  \caption{(a) upper panel: Time evolution of volume, energy, enthalpy and intensity of selected Bragg peaks after amorphization of Ge$_{136}$ at $p=100$ kbar and $T = 300$ K. (b) lower panel: Structural evolution of the system during the recrystallization of the amorphous phase. The figure was prepared using the OVITO package\cite{ovito}.}
 \label{fig:recrystallization_100kbar}
\end{figure}

Performing metadynamics further after the collapse of the clathrate is not justified since for the non-crystalline disordered phase the supercell edges do not anymore represent good collective variables and unphysical changes of the shape of the system might occur. Therefore, in order to further follow the evolution of the disordered system we switched off the Gaussians after metastep 83 and performed plain NPT MD for another 20000 steps (40 ps). 
Surprisingly, we observed a recrystallization of the disordered phase. Such ultrafast recrystallization observed in a short ab initio MD run at room temperature appears rather unusual. 
The evolution of the volume, energy, enthalpy and intensity of selected Bragg peaks is shown in Fig.\ref{fig:recrystallization_100kbar} (a). 
During the first 12000 steps the energy slowly decreases while some Bragg peaks grow slowly. Afterwards the peaks start to grow very fast, energy drops and a defective $\beta$-tin phase is formed. After 17000 steps the latter phase transforms into defective simple hexagonal phase. This last transformation is likely to be related to the presence of defects since the ideal sh phase becomes thermodynamically stable only at a higher pressure (see Fig.1 (SM)). We note that one cannot expect a formation of a perfect crystal since the number of atoms in the supercell is 136 which does not allow creation of perfect supercell with edges being small integer multiples of unit cell vectors of crystalline phases. The structural evolution of the system and formation of crystalline order from the amorphous phase is shown in Fig.\ref{fig:recrystallization_100kbar} (b) and Fig.4 (SM).

We note that the observed pressure-induced amorphization of the clathrate has features analogous to the "cold melting" suggested to operate, e.g., in pressure-induced amorphization of water ice resulting in the high-density amorphous phase \cite{Ice_Mishima_1984,MACHON2014216}. The melting line of the cubic diamond structure has a negative Clapeyron slope and the melting temperature reaches a minimum of 800 K at pressure of 10 GPa\cite{Ge_phase_diagram_2021}. In lack of a mechanism allowing a direct transformation of the clathrate to another crystalline phase, the existence of a low-free-energy supercooled liquid offers a kinetic route towards a disordered structure available upon compression of the clathrate, providing a rationale for the pressure-induced amorphization. 
Within this scenario the disordered phase prepared by pressure-induced amorphization of the clathrate should have the same structure and exhibit the same ultrafast recrystallization behaviour as the supercooled liquid obtained by fast cooling of liquid at the same pressure. 
We note that a similar fast recrystallization was recently described in very-high-density-amorphous (VHDA) Si\cite{Si_Nature2021}. It was not, however, observed in computational studies of Ge which focused on the LDA/HDA transition \cite{PhysRevB.66.041201,Mancini_2015}.

\begin{figure}[htpb]
  \includegraphics*[width=\columnwidth]{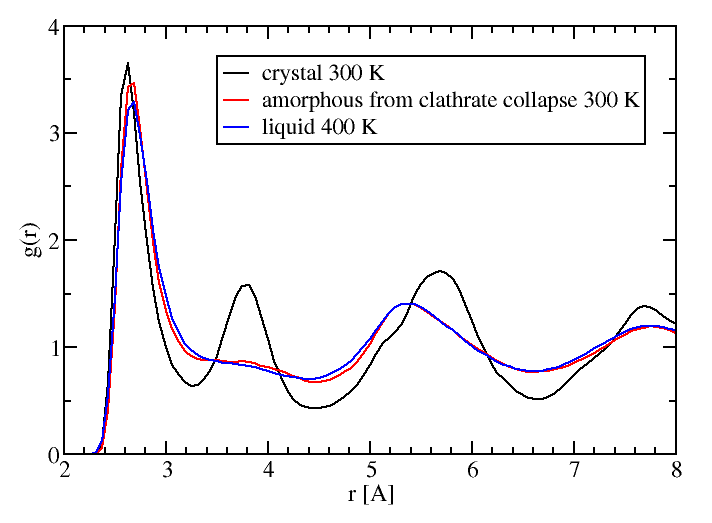}
  \caption{Radial distribution function of the system in disordered and crystalline state at various temperatures.}
 \label{fig:rdf}
\end{figure}

\begin{figure}[htpb]
  \includegraphics*[width=\columnwidth]{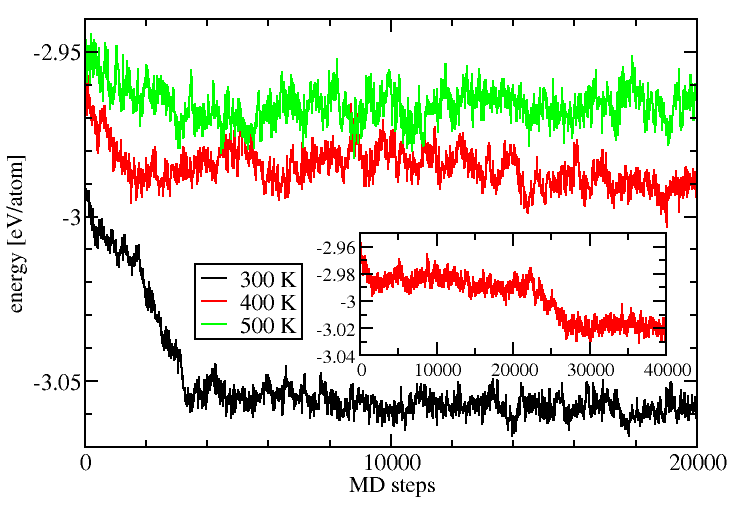}
  \includegraphics*[width=\columnwidth]{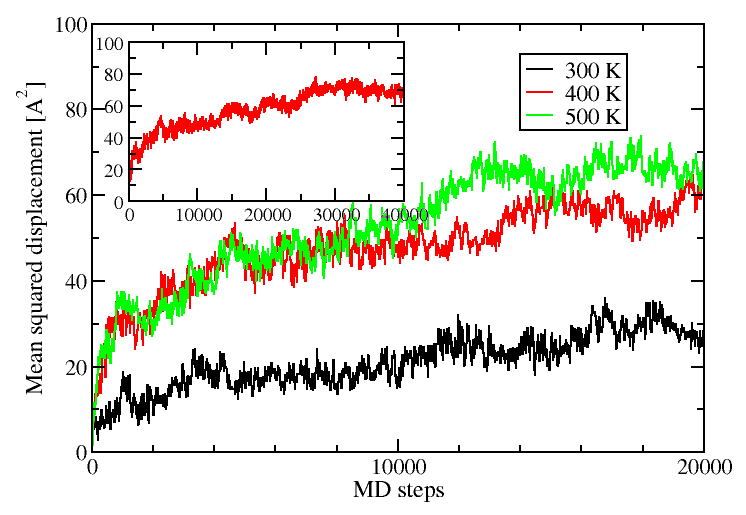}
  \caption{Energy evolution (a) (upper panel) and mean squared displacement of atoms (b) (lower panel) during liquid cooling over 20000 MD steps at $T=500, 400, 300$K. Crystallization starts at $T=300$ K after 4000 steps. Inset shows an extended run over total of 40000 MD steps at $T=400$ K where the system also crystallized after 27000 steps.}
   \label{fig:liquid_cooling}
\end{figure}

To test this hypothesis we 
melted the final recrystallized structure by heating it to 1000 K at 10 GPa over 20 ps. The liquid phase was subsequently quickly cooled down to 800 to 600 to 500 to 400 to 300 K, performing at each temperature a 20 - 40 ps MD run. The liquid phase persisted down to 300 K where the system again rapidly recrystallized within 20 ps as can be seen on energy evolution in Fig.~\ref{fig:liquid_cooling}(a), very similarly to the crystallization after clathrate collapse. 
The same crystallization occurred also at $T=400$ K within an extended run of additional 40 ps (inset of Fig.~\ref{fig:liquid_cooling}(a)).
Interestingly, also the radial distribution function of the amorphous phase from clathrate collapse at 300 K is very similar to that of the liquid before crystallization at 400 K (Fig.~\ref{fig:rdf}). A remarkable feature of both RDFs is the lack of the peak corresponding to the 2nd coordination shell of the crystal at 3.8 $\AA$. 
The pressure-induced collapse of the clathrate structure therefore produces an amorphous phase which is structurally very similar to the one obtained by fast liquid cooling. Importantly, both structures appear equally prone to ultrafast crystallization. We note that in order to crystallize quickly the supercooled liquid must still have sufficient diffusivity. We therefore monitored the mean squared displacement of the atoms at different temperatures which is shown in Fig.\ref{fig:liquid_cooling}(b). It can be seen that the liquid even when cooled down to 300 K does not represent a frozen glass where atoms merely vibrate around their average positions. The glass transition temperature must therefore be rather low, allowing the system to crystallize before it freezes to amorphous solid. This appears to be a rather unusual property of the supercooled liquid Ge under pressure.

The observed ultrafast recrystallization points to the lack of barrier separating the disordered and the crystalline phase. The crystallization in a strongly supercooled yet diffusive liquid thus occurs due to the loss of stability of the supercooled liquid. The whole process of transformation of the clathrate to the $\beta$-tin therefore represents a sequence of two off-equilibrium processes driven by the loss of stability of the parent phase. The initial pressure-induced amorphization of clathrate is triggered by high pressure while the subsequent fast recrystallization to $\beta$-tin is driven by low temperature. 

\begin{figure}[htpb]
 \includegraphics*[width=\columnwidth]{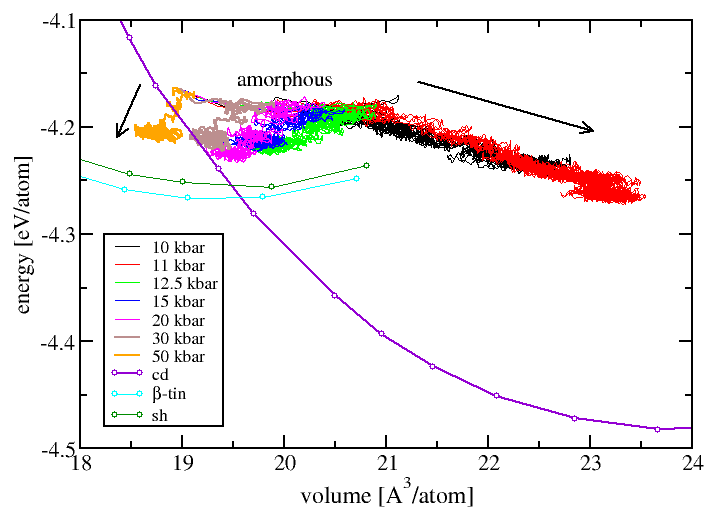}
 \includegraphics*[width=\columnwidth]{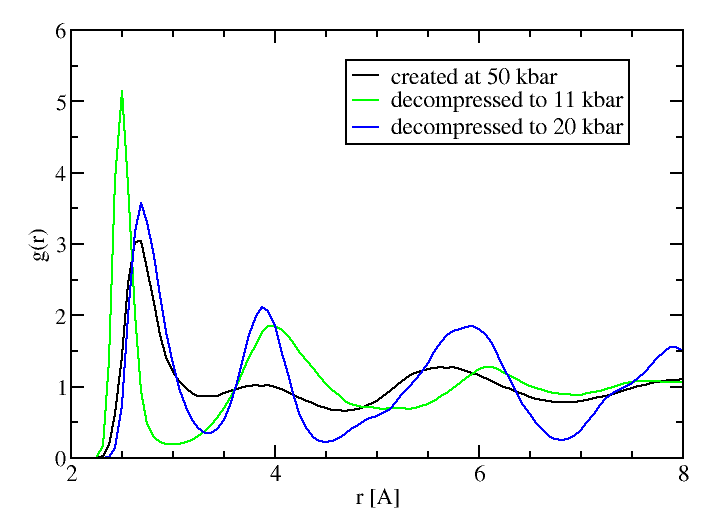}
  \caption{(a) upper panel: Energy vs. volume after sudden decompression of amorphous Ge prepared by compression of the clathrate Ge$_{136}$ at 50 kbar and 300 K. Arrows show the different direction of the evolution in the two regimes. (b) lower panel: Radial distribution function of the amorphous phase with incipient crystalline order at 50 kbar as well as of the phases after decompression to 20 and 11 kbar, showing different structural evolution.}
 \label{fig:decompression}
\end{figure}

Finally, we studied whether the ultrafast recrystallization of the dense amorphous phase persists also to lower pressures. To this end we repeated metadynamics simulation of the clathrate at a lower pressure of 5 GPa where the amorphization occurred after 278 metasteps. The resulting amorphous phase already exhibited an incipient crystalline order as can be seen on the RDF in Fig.\ref{fig:decompression}(b) where a small peak appears at the position of the 2$^{nd}$ coordination shell at 3.9 \AA. Afterwards we followed the structural evolution at even lower pressures, suddenly decompressing the phase from 50 kbar directly to pressures of 30, 20, 15, 12.5, 11 and 10 kbar. Interestingly, we found a boundary between two manifestly different regimes. The evolution of the system in the E-V plane is shown in Fig.\ref{fig:decompression} (a). It is seen that at pressures down to 12.5 kbar the system still approaches the E-V curves of the $\beta$-tin and sh phases (and eventually crystallizes). For decompression to 20 kbar this is demonstrated also in the Fig.\ref{fig:decompression} (b) where the initially small peak at 3.9 $\AA$ grows to a full peak of the crystalline phase while the volume shrinks (Fig.\ref{fig:decompression}(a)). On the other hand, the evolution of the RDF after decompression to 11 kbar is very different - the system strongly expands its volume (Fig.\ref{fig:decompression}(a)) and RDF evolves towards different peaks, where notably the first peak moves towards shorter bond. The different evolution pattern is also very clearly seen on the time evolution of the ADF shown for both cases in Fig.5 (SM).
In this regime, interestingly, the sudden expansion of the phase with incipient crystalline order drives the system to a disordered state followed by structural arrest, resulting in the tetrahedral amorphous LDA phase.
We note that the energy of the amorphous phase created by clathrate amorphization is only slightly above ($\sim 0.07$ eV) that of the $\beta$-tin and sh crystalline phases, indicating that the energy landscape of the relevant megabasin is rather flat (Fig.\ref{fig:decompression}(a)). This is also consistent with the lower barriers towards crystallization. On the other hand, the energy of the LDA phase at lower pressures remains much higher ($\sim 0.2$ eV) above the cubic diamond phase, consistent with much more structured energy landscape and a higher barrier making recrystallization much more difficult.

The above observations suggest that the potential energy landscape (PES) of the disordered/liquid HDA  phase is rather flat (unlike that of the low-pressure tetrahedral LDA Ge). This is compatible with the observed structural properties, namely nearly structureless RDF from 3 to 5 $\AA$ and broad distribution of the bond angle. This lack of roughness of the PES is also consistent with the low glass transition temperature, implying non-vanishing diffusivity even at low temperatures which eventually makes the ultrafast crystalization possible.

To conclude, we showed that clathrate Ge$_{136}$ upon compression at room temperature transforms to the $\beta$-tin phase, via a disordered amorphous phase. It thus provides an explicit example of structural transformation between two crystalline phases proceeding via non-crystalline intermediate. To further understand this interesting phenomenon it would be interesting to perform the experiment at different temperatures. While the pressure-induced amorphization can be expected to occur at all temperatures, the recrystallization is likely to be temperature dependent. Our prediction for the final phase is the following:
at $T > T_m$ (800 K) compression will melt clathrate, at $T_g < T < T_m$ compression will result in $\beta$-tin recrystalized from supercooled liquid and finally at sufficiently low temperature $T < T_g$ compression will create a metastable HDA amorphous phase.

\begin{acknowledgments}
M.R. and R.M. were supported by the VEGA project No.~1/0640/20 and by the Slovak Research and Development Agency under Contract No.~APVV-19-0371. The calculations were performed in the Computing Centre of the Slovak
Academy of Sciences using the supercomputing infrastructure acquired in project ITMS 26230120002 and 26210120002 (Slovak infrastructure for high-performance computing) supported by the Research \& Development Operational Programme funded by the ERDF. Part of calculations was perfomed on the GPU TITAN V provided by the NVIDIA grant. S.L. thanks the Leverhulme Trust for support under Project No. RPG-2020-052, as well as ARCCA Cardiff for computational support.\end{acknowledgments}

\end{document}